\documentclass[review]{elsarticle}

\usepackage{lineno,hyperref}
\usepackage{subfigure,graphicx}
\usepackage{amsmath,amsfonts,latexsym,amssymb,euscript,xr,setspace,mathrsfs,amsthm,booktabs,float,enumitem,diagbox,bm,comment}
\usepackage{tikz}
\usetikzlibrary{automata, positioning, arrows}
\tikzset{
->,
>=stealth',
node distance=3cm, 
every state/.style={thick, fill=gray!10},
initial text=$ $,
}

\theoremstyle{definition}
\newtheorem{definition}{Definition}[section]
\newtheorem{remark}{Remark}[definition]
\newtheorem{example}{Example}[section]

\begin{document}

\begin{frontmatter}

\title{Automaton of molecular perceptions in biochemical reactions}

\author{Stefano Maestri}\corref{mycorrespondingauthor}
\ead{stefano.maestri@unicam.it}
\author{Emanuela Merelli}

\address{School of Science and Technology, University of Camerino, Camerino, 62032, Italy}

\cortext[mycorrespondingauthor]{Corresponding author}

\begin{abstract}
Local interactions among biomolecules, and the role played by their environment, have gained increasing attention in modelling biochemical reactions. By defining the \textit{automaton of molecular perceptions}, we explore an agent-based representation of the behaviour of biomolecules in living cells. Our approach considers the capability of a molecule to perceive its surroundings a key property of bimolecular interactions, which we investigate from a theoretical perspective. Graph-based reaction systems are then leveraged to abstract enzyme regulation as a result of the influence exerted by the environment on a catalysed reaction. By combining these methods, we aim at overcoming some limitations of current kinetic models, which do not take into account local molecular interactions and the way they are affected by the reaction environment.\end{abstract}


\end{frontmatter}

\section{Introduction}
\label{sec:background}

Biochemical reactions, especially those involved in metabolic and signaling pathways, have been widely studied through computational approaches~\citep{berryMonte2002,takahashiSpace2005,bartocciDetecting2010,butiBioShape2010}. In particular, Systems Biology provides computational models focused mostly on the kinetic properties of the enzymatic reactions, represented through sets of differential equations~\citep{westerhoffSilicon2011b}; this approach may constitute a limitation, especially since kinetic parameters sampled \textit{in vitro} might not capture the dynamic interactions that enzymes carry out with their environment~\citep{teusinkCan2000}. In this context, the latter is also considered as homogeneous, and  enzymes as functions from reactants to products, overlooking the granularity that characterises molecular interactions. 

In recent years, this problem has been tackled by modelling and simulating the molecules involved in a biochemical reaction as \textit{agents}, autonomous systems able to perceive and interact with the other agents populating their environment~\citep{cannataAgentBased2013,merelliAgents2006}. Such an approach allows capturing the fundamental role that molecular perception has on the efficiency of a biochemical reaction, a property that has been experimentally observed in the form synchronised oscillations among interacting molecules~\cite{nardecchiaDetection2017a}.

According to the Michaelis-Menten model of enzyme kinetics, an enzymatic reaction can be represented as:
\begin{equation}
E + S \leftrightharpoons ES \rightarrow E + P
\end{equation}
where \textit{E} is an enzyme, \textit{S} its substrate and \textit{P} the product of the reaction catalysed by \textit{E}; assuming the steady-state approximation, we can consider ES as constant~\cite{johnsonOriginal2011,briggsNote1925}. 

As part of an agent-based model of a biochemical reaction, we described this pattern through a basic \textit{reaction automaton}~\citep{piangerelli2020visualising}. It was intended just to support the interaction phases accounted in the agent-based model and does not explicitly consider agent's perception, as well as ignores the effects of enzyme regulation on the generation of the reaction products.

In this manuscript, we refine this idea from a theoretical perspective, by considering the environment as a first-class component of the system, which is perceived by the agents representing the molecule involved in the reaction; we also leverage the capabilities of the \textit{graph-based reaction systems}~\citep{kreowskiGraph2019a} to move a first step in modelling the regulation carried out, on the enzymatic activity, by specific molecules of the environment.

\section{Materials and Methods}
\label{sec:methods}
The work described in this manuscript relies on automata theory, agent-based modelling and graph-based reaction systems. In this section, we provide a brief introduction to graph-based reaction systems, which, among the other, is a relatively new theory, and maybe less known.

Reaction systems were introduced as a formal framework to model biochemical reactions in living cells. Although the original approach is purely set-theoretic~\citep{ehrenfeuchtReaction2007}, a graph-based description is provided by Kreowski and Rozenberg~\citep{kreowskiGraph2019a}, allowing to deal with graph-related problems, such as those we tackle in this manuscript. We report here some basic concepts, as described in the aforementioned source article, needed for the theoretical constructions proposed in the next section.

\medskip
According to Kreowski and Rozenberg~\citep{kreowskiGraph2019a}:

\begin{itemize}
\item A \textit{directed and edge-labelled graph} is a system $G = (V ,\Sigma, E)$, where $V$ is a finite set of nodes, $\Sigma$ is a finite set of edge labels, and $E \subseteq V \times V \times \Sigma$ is a set of edges.
\item Given the graphs $H$ and $G$, such that $\Sigma_{H} = \Sigma_{G}$, $H$ is a \textit{subgraph} of $G$ if $V_{H} \subseteq V_{G}$, and $E_{H} \subseteq E_{G}$. We denote with $Sub(G)$ the \textit{set of all the subgraphs of} $G$.
\item A \textit{selector} $\mathcal{S}$ of a graph $G = (V,E)$, is an ordered pair $\mathcal{S} = (X,Y)$ such that $X \subseteq V$ and $Y \subseteq E$. 
\item The \textit{extraction} $U(H)$ of $H \in Sub(G)$ is the pair $U(H) = (V_{H}, E_{H})$, respectively of the set of nodes and set of edges of the graph $H$.
\item A  \textit{reaction b} over a non-empty, directed, and edge-labelled graph $B = (V_{B}, \Sigma_{B}, E_{B})$, called the \textit{background graph}, is a triple $b = (R, I, P)$, where $R$ and $P$ are non-empty subgraphs of $B$, and $I$ is a selector of $B$ such that $I \cap \,U(R) = (\emptyset,\emptyset)$.
\item Let $T$ be a subgraph of $B$.
\begin{itemize}
\item A reaction $b = (R, I, P)$ over $B$ is \textit{enabled by} $T$, denoted by $en_{b}(T)$, if $R \in Sub(T)$ and $I \cap U(T) =  (\emptyset,\emptyset)$.
\item The \textit{result of a reaction} $b$ on $T$, denoted by $res_{b}(T)$, is defined by $res_{b}(T) = P_{b}$ if $en_{b}(T)$; $res_{b}(T) = \emptyset$ otherwise.
\item The \textit{result of a set of reactions} $A$ over $B$ on $T$ , denoted by $res_{A}(T)$, is defined by: $res_{A}(T) = \bigcup_{b \in A} res_{b}(T)$.
\item A \textit{graph-based reaction system} is a pair $\mathcal{A} = (B, A)$, where $B$ is a finite non-empty graph, and $A$ is a set of reactions over $B$.
Also, $\mathcal{A}$ induces the function $res_{\mathcal{A}} : Sub(B) \rightarrow Sub(B)$, called the \textit{result function} of $\mathcal{A}$, such that, for each state $T \in Sub(B)$, $res_{\mathcal{A}}(T) = res_{A}(T)$.
\end{itemize}
\end{itemize}

\section{Results}
\label{sec:results}

\subsection{A basic enzymatic reaction model}
An enzymatic reaction, as introduced in Section~\ref{sec:background}, can be modelled by explicitly taking into account the possible role of coenzymes, such as ATP or NADH. 

Let $\Sigma = \{c,p,r,s\}$ be an alphabet, where $c$ represents a coenzyme, $p$ the main product of the reaction, $r$ the result of the conversion of a coenzyme (if any), and $s$ a substrate molecule. We clarify that a coenzyme is, indeed, substrate of the related enzyme, but we use the above naming to emphasise its specific role of supporting the reaction. Let also $Q = \{E, Ec, Ep, Er, Es, Esc\}$ be a set of states, where $E$ represents a free enzyme, $Ec$ the binding of a coenzyme, $Es$ the enzyme bound to a substrate molecule, $Esc$ an enzyme saturated by a substrate molecule and a coenzyme; $Ep$ and $Er$ are the states through which, respectively, the main product of the reaction and the product of the coenzyme conversion are released. Binding a substrate molecule and a coenzyme can happen without a specific order, however there is no case in which a coenzyme can directly generate a product. We also assume that, when the reaction has two products, they are not released exactly in the same instant, even though, from the biological perspective, this time delay may be negligible. The basic behaviour of an enzymatic reaction can thus be described by the deterministic finite state automaton $\mathcal{R} = (Q,\Sigma,\delta, s_{0}, \{f\})$, with $s_{0} = f = E$ being the initial and final state, and $\delta: Q \times \Sigma \rightarrow Q$  the state transition relation defined as follows:

\begin{center}
\begin{tabular}{| l | c | c | c | c |} 
 \toprule
 \diagbox{\footnotesize \bf State}{\footnotesize \bf Input}& $\bm s$ & $\bm c$ & $\bm p$ & $\bm r$ \\ 
 \midrule
 $\bm E$ & $Es$  & $Ec$  & $-$ & $-$\\ 
 \hline
 $\bm{Es}$ & $-$  & $Esc$  & $E$ & $-$\\  
 \hline
 $\bm{Ec}$ & $Esc$  & $-$  & $-$ & $-$\\  
 \hline
 $\bm{Esc}$ & $-$  & $-$  & $Er$ & $Ep$\\  
 \hline
 $\bm{Ep}$ & $-$  & $-$  & $E$ & $-$\\  
 \hline
 $\bm{Er}$ & $-$  & $-$  & $-$ & $E$\\ 
 \bottomrule
\end{tabular}
\end{center}
\label{def:atuoma}

Note that, from $Es$ (i.e., the state in which the enzyme $E$ is bound to the substrate molecule \textit{s}), the automaton can reach either the state $E$, by generating the product $p$, or the state $Esc$, by binding the coenzyme $c$. This approach allows the enzymatic reaction automaton to generalise the behaviour of enzymes that require a coenzyme to catalyse the reaction, as well as of those able to directly facilitate the substrate-to-product conversion. 

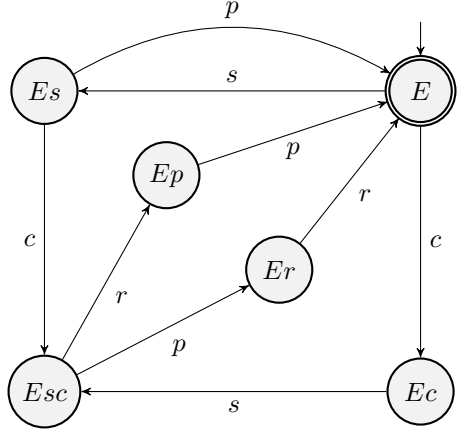
\begin{figure}[H]
\centering
\begin{tikzpicture}
\node[state, initial, initial where= above, accepting] (q1) {$E$};
\node[state, yshift=-1cm, below of=q1] (q2) {$Ec$};
\node[state, xshift = -2cm, left of=q1] (q3) {$Es$};
\node[state, yshift=-1cm, below of=q3] (q4) {$Esc$};
\node[state, yshift=1cm, xshift=-0.5cm, below right of=q3] (q5) {$Ep$};
\node[state, yshift=-0.5cm, xshift=1cm, above right of=q4] (q6) {$Er$};
\draw
(q1) edge[right] node{$c$} (q2)
(q1) edge[above] node{$s$} (q3)
(q2) edge[below] node{$s$} (q4)
(q3) edge[left] node{$c$} (q4)
(q4) edge[below right] node{$r$} (q5)
(q4) edge[below right] node{$p$} (q6)
(q5) edge[below] node{$p$} (q1)
(q6) edge[below right] node{$r$} (q1)
(q3) edge[bend left, above] node{$p$} (q1);

\end{tikzpicture}
\caption{Automaton representing the basic behaviour of an enzymatic reaction, which may or may not involve a coenzyme.}
\label{fig:simpaut}
\end{figure}

The automaton state graph is shown in Fig.~\ref{fig:simpaut}.

This model of enzymatic reaction is a refined version of the one described in~\citep{maestriAlgebraic2020a}, and does not consider the effect of molecular perception. However, form an agent-based perspective, enzymes are active entities able to perceive and modify their environment, which is populated by agents modelling substrate molecules and other enzymes.

\subsection{\bf \it Enhancing the enzymatic reaction model with perception}

To endow $\mathcal{R}$ with perception, we should add a level of nondeterminism, and introduce $\epsilon$-transitions. The latter are needed to model the unstable states in which an enzyme perceives a cognate molecule, but the actual binding is not yet certain.

We therefore construct the extended  reaction automaton $\mathcal{R^{'}}$, which differs from the previous one in its set of states and transition function.

\begin{definition}[Perception-based reaction automaton] We define \textit{perception-based reaction automaton} the 5-tuple $\mathcal{R^{'}} = (Q^{'},\Sigma \cup \{\epsilon\},\delta^{'},s_{0},\{f\})$, such that:\begin{itemize}
\item $Q^{'} = \mathscr{S} U \mathscr{P}$, where $\mathscr{S} = \{E,Ec,Ep,Er,Es,ES\}$ and $\mathscr{P} = \{E_{c},Ec_{s},E_{s}, Es_{c}\}$.
\begin{itemize}
\item $\mathscr{S}$ is the set of \textit{stable states}, where an enzyme $E$ is free, bound to a cognate molecule or coenzyme in a molecular complex ($Es$, $Ec$), or is saturated ($ES$) and therefore ready to catalyse the reaction, release its product (through the $Er$ and $Ep$ states, when a coenzyme is involved), and thus return free;  
\item $\mathscr{P}$ represents the set of \textit{perceiving states}, where the free enzyme (or a complex) perceives a cognate molecule $s$ or coenzyme $c$ (indicated as subscripts).
\end{itemize}
\item $\Sigma = \{c,p,r,s\}$ and  $s_{0} = f = E$
\item $\delta^{'}: Q \times \Sigma \cup \{\epsilon\} \rightarrow 2^{Q}$, is given by the following transition table:
\end{itemize}

\begin{center}
\begin{tabular}{| l | c | c | c | c | c |} 
 \toprule
 \diagbox{\footnotesize \bf State}{\footnotesize \bf Input}& $\bm s$ & $\bm c$ & $\bm p$ & $\bm r$ & $\bm \epsilon$ \\ 
 \midrule
 $\bm E$ & $\{E_{s}\}$  & $\{E_{c}\}$  & $\{\}$ & $\{\}$ & $\{E\}$\\ 
 \hline
 $\bm{Es}$ & $\{\}$  & $\{Es_{c}\}$  & $\{\}$ & $\{\}$ & $\{Es\}$\\  
 \hline
 $\bm{Ec}$ & $\{Ec_{s}\}$  & $\{\}$  & $\{\}$ & $\{\}$ & $\{Ec\}$\\  
 \hline
 $\bm{ES}$ & $\{\}$  & $\{\}$  & $\{Er,E\}$ & $\{Ep\}$ & $\{\}$\\  
 \hline
 $\bm{Ep}$ & $\{\}$  & $\{\}$  & $\{E\}$ & $\{\}$ & $\{\}$\\  
 \hline
 $\bm{Er}$ & $\{\}$  & $\{\}$  & $\{\}$ & $\{E\}$ & $\{\}$\\  
 \hline
 $\bm{E_{s}}$ & $\{Es,ES\}$  & $\{\}$  & $\{\}$ & $\{\}$ & $\{E\}$\\  
 \hline
 $\bm{Es_{c}}$ & $\{\}$  & $\{ES\}$  & $\{\}$ & $\{\}$ & $\{Es\}$\\  
 \hline
 $\bm{E_{c}}$ & $\{\}$  & $\{Ec\}$  & $\{\}$ & $\{\}$ & $\{E\}$\\ 
 \hline
 $\bm{Ec_{s}}$ & $\{ES\}$  & $\{\}$  & $\{\}$ & $\{\}$ & $\{Ec\}$\\  
 \bottomrule
\end{tabular}
\end{center}
\label{def:pbra}
\end{definition}

As mentioned above, the $\epsilon$-transition allows a stable enzyme or complex to not perceive any cognate molecule in its surrounding environment or to return, if the binding of the perceived molecule does not happen, from a \textit{perceiving state} to a previous \textit{stable state}. 

Introducing \textit{perceiving states} in the model forces the  nondeterminism on the pairs $(E_{s},s)$ and $(ES,p)$ to distinguish the case in which the reaction involves a coenzyme from that where the sole substrate molecule $s$ is converted. 

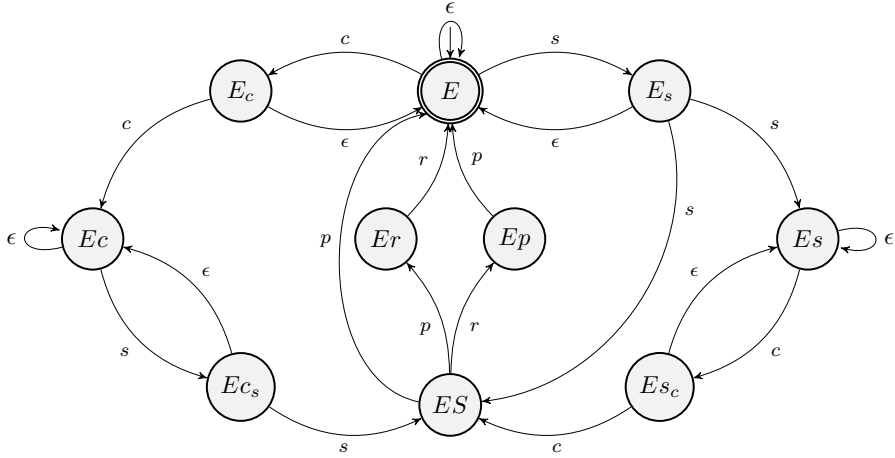
\begin{figure}[H]
\resizebox{\textwidth}{!}{
\begin{tikzpicture}
\node[state, initial,  initial where= above, accepting] (q1) {$E$};
\node[state, right of=q1] (q2) {$E_{s}$};
\node[state, left of=q1] (q3) {$E_{c}$};
\node[state, below left of=q3] (q4) {$Ec$};
\node[state, below right of=q2] (q5) {$Es$};
\node[state, below right of=q4] (q6) {$Ec_{s}$};
\node[state, below left of=q5] (q7) {$Es_{c}$};
\node[state, below of=q1, yshift=-1.5cm] (q8) {$ES$};
\node[state, below left of=q1, xshift=1.2cm] (q9) {$Er$};
\node[state, below right of=q1, xshift=-1.2cm] (q10) {$Ep$};
\draw
(q1) edge[above, bend left] node{\footnotesize $s$} (q2)
(q1) edge[above, bend right] node{\footnotesize $c$} (q3)
(q3) edge[below, bend right] node{\footnotesize $\epsilon$} (q1)
(q2) edge[below, bend left] node{\footnotesize $\epsilon$} (q1)
(q3) edge[above left, bend right] node{\footnotesize $c$} (q4)
(q2) edge[above right, bend left] node{\footnotesize $s$} (q5)
(q4) edge[loop left] node{$\epsilon$} (q4)
(q5) edge[loop right] node{$\epsilon$} (q5)
(q4) edge[below left, bend right] node{\footnotesize $s$} (q6)
(q5) edge[below right, bend left] node{\footnotesize $c$} (q7)
(q6) edge[above right, bend right] node{\footnotesize $\epsilon$} (q4)
(q7) edge[above left, bend left] node{\footnotesize $\epsilon$} (q5)
(q7) edge[below, bend left] node{\footnotesize $c$} (q8)
(q6) edge[below, bend right] node{\footnotesize $s$} (q8)
(q8) edge[anchor=south, below left, bend right = 20] node{\footnotesize $p$} (q9)
(q8) edge[below right, bend left = 20] node{\footnotesize $r$} (q10)
(q9) edge[above left, bend right = 20] node{\footnotesize $r$} (q1)
(q10) edge[above right, bend left = 20] node{\footnotesize $p$} (q1)
(q2) edge[pos=0.2,below right, bend left = 50] node{\footnotesize $s$} (q8)
(q8) edge[above left, bend left = 80] node{\footnotesize $p$} (q1.225)
(q1) edge[loop above] node{$\epsilon$} (q1);
\end{tikzpicture}}
\caption{State graph of the perception-based reaction automaton; it represents the generalised behaviour of the perception-based enzymatic reactions (see Definition~\ref{def:per}).}
\label{fig:percgraph}
\end{figure}

The $ES$ state represents a saturated enzyme, which is not able to perceive its surroundings; similarly, the enzyme has no perception in the $Er$ and $Ep$ states, where it releases the products of the reaction. This behaviour is represented by the state graph in Fig.~\ref{fig:percgraph}.

\begin{definition}[Enzyme perception and reaction anabler]
Given a perception-based reaction automaton $\mathcal{R^{'}} = (Q^{'},\Sigma \cup \{\epsilon\},\delta^{'},s_{0},\{f\})$, and a set $\Pi \subset \Sigma/\{p\}$, the function $\mu : \mathscr{S} \times \Pi \rightarrow \mathscr{P}$ is called \textit{enzyme perception in} $\mathcal{R^{'}}$.  The set $\Pi$, defined \textit{reaction enabler}, contains all the cognate molecules that the enzyme must perceive in its surrounding environment in order to proceed during the reaction.
\label{def:pi}
\end{definition}

It is possible to identify an \textit{equivalence relation} $\equiv$ that characterises the \textit{equivalence class}  $[\mathcal{R^{'}}]_{\equiv}$ of the automaton $\mathcal{R^{'}}$. Although a rigorous definition of this class is beyond the scope of the present article (and will be provided in a future work), intuitively, we can say that it represents the class of all the automata that show the same behaviour of $\mathcal{R^{'}}$.

\begin{definition}[Perception-based enzymatic reaction]
\label{def:per}
Being $[\mathcal{R^{'}}]_{\equiv}$ the equivalence class of the perception-based automaton $\mathcal{R^{'}}$, a 5-tuple $\mathcal{R}_{p} = (Q,\Sigma,\delta,s_{0},F)$ is a \textit{perception-based enzymatic reaction} iff $\mathcal{R}_{p} \in [\mathcal{R^{'}}]_{\equiv}$ (that is, $\mathcal{R}_{p} \equiv \mathcal{R^{'}}$).
\end{definition}

\begin{example}
The phosphorylation of \texttt{fructose 6-hosphate} (\texttt{F6P}) to \texttt{fructose 1,6-bisphosphate} (\texttt{F16bP}), which is carried out by the \texttt{phosphofructokinase} (\texttt{PFK}) and coupled with the hydrolysis of \texttt{ATP} to \texttt{ADP}, is modelled through the \textit{perception-based enzymatic reaction} shown in Fig.~\ref{fig:percaut}.

\begin{figure}[H]
\resizebox{\textwidth}{!}{
\begin{tikzpicture}
\node[state, initial,  initial where= above, accepting] (q1) {$PFK$};
\node[state, right of=q1] (q2) {$PFK_{F6P}$};
\node[state, left of=q1] (q3) {$PFK_{ATP}$};
\node[state, below left of=q3] (q4) {$PFK$\footnotesize$ATP$};
\node[state, below right of=q2] (q5) {$PFK$\footnotesize$F6P$};
\node[state, below of=q4] (q6) {$PFK$\footnotesize$ATP$$_{F6P}$};
\node[state, below of=q5] (q7) {$PFK$\footnotesize$F6P$$_{ATP}$};
\node[state, below of=q1, yshift=-4cm] (q8) {$PFK$\footnotesize$ATPF6P$};
\node[state, below left of=q1, yshift=-1cm] (q9) {$PFK$\footnotesize$ADP$};
\node[state, below right of=q1,yshift=-1cm] (q10) {$PFK$\footnotesize$F16bP$};
\draw
(q1) edge[above, bend left] node{\footnotesize $F6P$} (q2)
(q1) edge[above, bend right] node{\footnotesize $ATP$} (q3)
(q3) edge[below, bend right] node{\footnotesize $\epsilon$} (q1)
(q2) edge[below, bend left] node{\footnotesize $\epsilon$} (q1)
(q3) edge[above left, bend right] node{\footnotesize $ATP$} (q4)
(q2) edge[above right, bend left] node{\footnotesize $F6P$} (q5)
(q4) edge[loop left] node{$\epsilon$} (q4)
(q5) edge[loop right] node{$\epsilon$} (q5)
(q4) edge[above left, bend right] node{\footnotesize $F6P$} (q6)
(q5) edge[above right, bend left] node{\footnotesize $ATP$} (q7)
(q6) edge[above right, bend right] node{\footnotesize $\epsilon$} (q4)
(q7) edge[above left, bend left] node{\footnotesize $\epsilon$} (q5)
(q7) edge[below, bend left] node{\footnotesize $ATP$} (q8)
(q6) edge[below, bend right] node{\footnotesize $F6P$} (q8)
(q8) edge[below left, bend right=20] node{\footnotesize $F16bP$} (q9)
(q8) edge[below right, bend left=20] node{\footnotesize $ADP$} (q10)
(q9) edge[above left, bend right=20] node{\footnotesize $ADP$} (q1)
(q10) edge[above right, bend left=20] node{\footnotesize $F16bP$} (q1)
(q1) edge[loop above] node{$\epsilon$} (q1);
\end{tikzpicture}}
\caption{Perception-based enzymatic reaction performed by \texttt{phosphofructokinase} (\texttt{PFK}).}
\label{fig:percaut}
\end{figure}
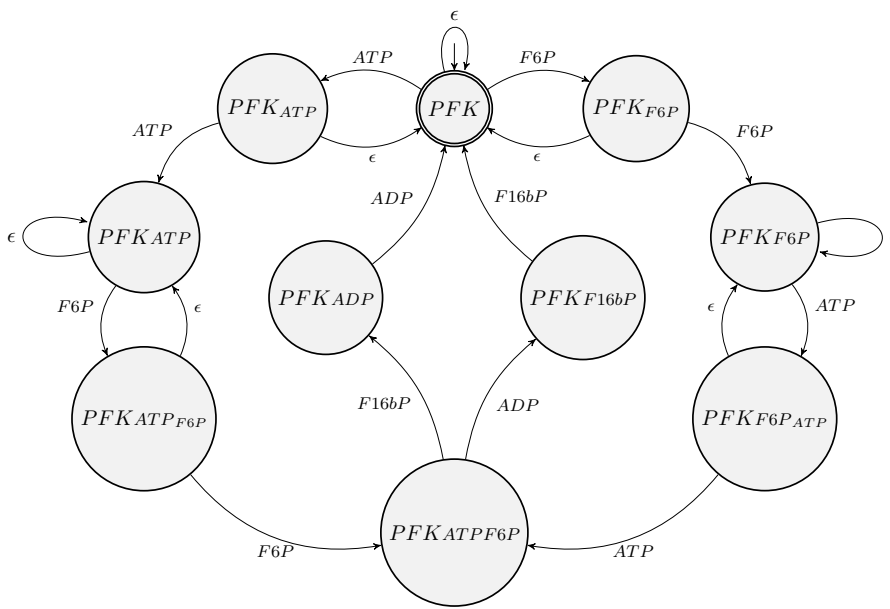
\end{example}

\subsection{\bf \it Enzyme regulation in perception-based reactions}

An important property of the enzymatic reactions, especially those involved in metabolic and signaling pathways, is the capability of being regulated (that is, activated or inhibited) in the process of generating their products. Enzyme regulation is often significantly complex; we propose here a first approach to tackle this property in modelling a perception-based enzymatic reaction. The core idea is to abstract the inhibition process as a ``switch'' that completely disable the reaction, so that it cannot generate its products; to implement this idea, we rely on the framework of \textit{graph-based reaction systems}, treated in~\citep{kreowskiGraph2019a}, and outlined in Section~\ref{sec:methods} of this manuscript.

According to~\citep{kreowskiGraph2019a}, a \textit{reaction} over a non-empty, directed and edge-labelled, graph $B = (V_{B}, \Sigma_{B}, E_{B})$, called \textit{background graph}, is a triple $b = (R, I, P)$, where $R$ and $P$ are non-empty subgraphs of $B$, and $I = (V_{I}, E_{I})$ is a selector of $B$ such that $I \cap \,U(R) = (\emptyset,\emptyset)$. $R$ is the \textit{reactant graph} of $b$, $I$ is the \textit{inhibitor} of $b$, and $P$ is the\textit{ product graph} of $b$.

We recall that, a \textit{selector} $\mathcal{S}$ of a graph $G = (V,E)$, is an ordered pair $\mathcal{S} = (X,Y)$ such that $X \subseteq V$ and $Y \subseteq E$, and that the \textit{extraction} $U(H)$ of $H \in Sub(G)$ is the pair $U(H) = (V_{H}, E_{H})$, respectively of the set of nodes and set of edges of the graph $H$.

\medskip
In our setting, the background graph $B$ represents all the possible spatial configurations that the molecules can assume in the modelled portion of cytoplasm. Each of them constitutes the \textit{environment} of the perception-based enzymatic reactions.

To introduce enzyme perception in a graph-based reaction, we need to define a new selector $\mathcal{S}_{\Pi}$ (perception selector), which identifies all the edges of $B$ labelled with the substrate molecules enabling a perception-based enzymatic reaction $\mathcal{R}_{b}$.

\begin{definition}[Perception selector]
Being $\mathcal{R}_{p} = (Q,\Sigma,\delta,s_{0},F)$ a perception-based enzymatic reaction, $B = (V_{B}, \Sigma_{B}, E_{B})$ a background graph, and $\Upsilon =\{I\,|\, I = (V_{I}, E_{I}) \in b, \forall$ $b = (R,I,P)$ over $B\}$, such that $\Sigma_{B} = \Sigma \cup \Gamma$, where $\Gamma = \{i \,|\, (v,v',i) \in E_{I}$ and $v,v' \in V_{I}, \forall\, I = (V_{I},E_{I}) \in \Upsilon\}$. Being also $\Pi \subset \Sigma$  the reaction enabler of $\mathcal{R}_{b}$, and $E_{\Pi} \subset E_{B}$ such that $E_{\Pi} =\{(v,v',x)\,|\,x \in \Pi\}$, a \textit{perception selector} of $\mathcal{R}_{p}$ is a pair $\mathcal{S}_{\Pi} = (\{v,v'\,|\,(v,v',x) \in E_{\Pi}\}, E_{\Pi}).$
\label{def:select}
\end{definition}

Now, we can extend the \textit{enabled reactions} defined in~\citep{kreowskiGraph2019a} (see Section~\ref{sec:methods}) with the concept of enzyme perception. 
\newpage

\begin{definition}[Perception-enabled reaction, result by perception]
\label{def:pen}
Given a graph $T \in Sub(B)$, a reaction $b = (R, I, P)$ over $B$ is \textit{perception-enabled} by $T$, denoted $pen_{b}(T)$, iff
\begin{itemize}
\item $R,P \in Sub(T)$ and $U(R) \cap \mathcal{S}_{\Pi} \neq (\emptyset,\emptyset)$; 
\item $U(T) \cap I = (\emptyset,\emptyset)$.
\end{itemize}

Otherwise, b is \textit{disabled} by $T$, denoted $dis_{b}(T)$.

The \textit{result by perception} $resp_{b}(T)$ of a reaction $b$ on $T$, is defined as $resp_{b}(T) = P_{b}$ iff $pen_{b}(T)$; if $dis_{b}(T)$, then $resp_{b}(T) = \emptyset_{\Sigma}$. The result by perception of a set of reactions $A$ over $B$ on $T$, denoted by $resp_{A}(T)$, is defined as $resp_{A}(T) = \bigcup_{b \in A} resp_{b}(T)$.
\label{def:enab}
\end{definition}

The subset $T$ of $B$ represents the configuration that the molecules in the modelled cytoplasm portion assume at a specific time (which may be, as an example, the time-step of an agent-based simulation). A perception-based reaction $\mathcal{R}_{b}$ is enabled by this  \textit{environment configuration} only if the enzyme that carries out  $\mathcal{R}_{b}$ perceives its cognate molecules (i.e., the substrate molecules that enable the reaction) and if it is not inhibited. In Definition~\ref{def:pen}, such conditions are guaranteed
ed by imposing, respectively, $U(R) \cap \mathcal{S}_{\Pi} \neq (\emptyset,\emptyset)$ and $U(T) \cap I = (\emptyset,\emptyset)$.

\begin{remark}
Since $R,P \in Sub(T) \in Sub (B)$ and $\mathcal{S}_{\Pi}$ is defined over $\Pi \subset \Sigma \subset \Sigma_{B}$, perception-enabled reactions are special cases of graph-based reactions. 
\label{lem:perc}
\end{remark}

Kreowski and Rozenberg~\citep{kreowskiGraph2019a} define a \textit{graph-based reaction system} as a pair $\mathcal{A} = (B, A)$, where $B$ is a finite non-empty graph, and $A$ is a set of reactions over $B$.

\begin{definition}[Result by perception function]
A function $resp_{\mathcal{A}} : Sub(B) \rightarrow Sub(B)$, called the \textit{result by perception function} of $\mathcal{A} = (B, A)$, is such that, for each state $T \in Sub(B)$, $resp_{\mathcal{A}}(T) = resp_{A}(T)$.
\label{def:rs}
\end{definition}

\begin{definition}[Perception-based enzymatic reaction inhibition]
A perception-based enzymatic reaction $\mathcal{R}_{p} = (Q,\Sigma,\delta,s_{0},F)$ is \textit{inhibited by a configuration of its environment}, defined through a graph T, iff the graph-based reaction system $\mathcal{A}(\mathcal{R}_{p}) = (B(\mathcal{R}_{p}), A(\mathcal{R}_{p}))$, associated to $\mathcal{R}_{p}$ through the steps described in Section 5 of~\citep{kreowskiGraph2019a}, is such that $T \in Sub((B(\mathcal{R}_{p}))$, and, $\forall$ reaction $b = (R,I,P) \in A(\mathcal{R}_{p})$, $\exists I = (V_{I},E_{I})$ such that $U(T) \cap I \neq (\emptyset,\emptyset)$, that is, $resp_{b}(T) = \emptyset_{\Sigma}$. 
\end{definition}

\section{Conclusions}
This manuscript provides a brief introduction on how finite-state automata and graph-based reaction systems can be combined to model the capabilities of biomolecules to perceive and interact with their environment. 

We start by refining a previously introduced reaction automaton~\cite{maestriAlgebraic2020a,piangerelli2020visualising}, and by endowing this model with \textit{environment perception}. Providing molecules with the ability to perceive cognate partners in their surroundings allows us to identify the equivalence class of the \textit{perception-based reaction automata}, to which we associate graph-based reaction systems~\cite{kreowskiGraph2019a}. From this new perspective, the environment is seen as a background graph $B$, while a molecules' space configuration (e.g., at a given time) as a subgraph $T$ that determines if a reaction can be carried out or not according to that specific context. Given these premises, we are then able to formally define the\textit{ enzymatic reaction inhibition}, carried out by the reaction environment, as a consequence of molecular perception.

Our analysis aims to set the basis for a novel modelling approach that, laying on solid formal frameworks, takes into account the fundamental effects of the reaction environment on biomolecular interactions. 

Although the proposed approach is strongly theoretical, it can be expanded in order to specify the behaviour of interacting molecules in a multiagent simulation~\cite{piangerelli2020visualising}. Further improvements of this work will also model the reversibility of perception-enabled reaction systems and introduce a way to influence the reaction direction through measurement values~\cite{bagossySimulating2020,ehrenfeuchtReaction2017}.
 
\section*{\bf Acknowledgments}
This article is the result of the research project funded by the Future and Emerging Technologies (FET) programme within the Seventh Framework Programme (FP7) for Research of the European Commission, under the FET-Proactive grant agreement TOPDRIM (\url{www.topdrim.eu}), number FP7-ICT- 318121.

\section*{Author contributions statement}

EM designed and supervised the research. SM implemented the method, performed the research and wrote the paper. All authors revised the paper and approved the final manuscript.

\section*{Competing interests}

The authors declare that they have no known competing financial interests or personal relationships that could have appeared to influence the work reported in this paper.

\bibliography{references}

\end{document}